\documentclass[aps,jmp,amsmath,amssymb,
preprint,%
]
{revtex4}

\usepackage{graphicx}
\usepackage{dcolumn}
\usepackage{bm}
\usepackage{hyperref}

\begin{document}

\title[]{How local is the Phantom Force?}

\author{Thorsten Wutscher}
\affiliation{Institute of Experimental and Applied Physics, University of Regensburg, 93040 Regensburg, Germany}

\author{Alfred J. Weymouth}
 \email{jay.weymouth@physik.uni-r.de}
\affiliation{Institute of Experimental and Applied Physics, University of Regensburg, 93040 Regensburg, Germany}

\author{Franz J. Giessibl}
\affiliation{Institute of Experimental and Applied Physics, University of Regensburg, 93040 Regensburg, Germany}

\date{\today}

\begin{abstract}
The phantom force is an apparently repulsive force, which can dominate the atomic contrast of an AFM image when a tunneling current is present. We described this effect with a simple resistive model, in which the tunneling current causes a voltage drop at the sample area underneath the probe tip. Because tunneling is a highly local process, the areal current density is quite high, which leads to an appreciable local voltage drop that in turn changes the electrostatic attraction between tip and sample. However, Si(111)-7$\times$7 has a metallic surface-state and it might be proposed that electrons should instead propagate along the surface-state, as through a thin metal film on a semiconducting surface, before propagating into the bulk. In this article, we investigate the role of the metallic surface-state on the phantom force. First, we show that the phantom force can be observed on H/Si(100), a surface without a metallic surface-state. Furthermore, we investigate the influence of the surface-state on our phantom force observations of Si(111)-7$\times$7 by analyzing the influence of the macroscopic tip radius $R$ on the strength of the phantom force, where a noticeable effect would be expected if the local voltage drop would reach extensions comparable to the tip radius. We conclude that a metallic surface-state does not suppress the phantom force, but that the local resistance $R_s$ has a strong effect on the magnitude of the phantom force.
\end{abstract}

\maketitle
\section{Introduction \label{sec:Introduction}}

Scanning probe microscopy (SPM) offers the possibility to determine structural and electronic properties of a surface on the atomic level \cite{Ternes2011,Sun2005}. The two most common SPM techniques are scanning tunneling microscopy (STM) and frequency-modulation atomic force microscopy (FM-AFM). With combined STM and FM-AFM, we recently observed a so-called phantom force on Si(111)-7$\times$7 \cite{Weymouth2011}. When the tip is too far from the surface to allow the resolution of chemical contrast in the force channel, atomic contrast can still be observed in constant-height mode with an applied bias. The resulting images appeared similar to the tunneling current images. In our proposed model, the tunneling current is injected from the tip into a small area within a radius of $\approx$\,100\,pm, the approximate atomic radius of Si. This causes
an appreciable voltage drop that decreases the electrostatic attraction between tip and sample as a function of tunneling current, causing these phantom force images. However, a highly localized voltage drop leading to the phantom force effect appears to be incompatible with the existence of the metallic surface-state of the Si(111)-7$\times$7 surface.\\
The Si(111)-7$\times$7 surface is described by the dimer-adatom-stacking-fault (DAS) model \cite{Takayanagi1985}. One
unit cell of the surface consists of 12 adatoms, which have partially filled dangling bonds, forming a metallic surface-state
\cite{Mauerer2006}. An intriguing question is how electrons propagate through the metallic surface-state, which is still under
discussion and we refer the reader to Refs.\cite{Persson1984, Hasegawa1992, Hasegawa1994, Heike1998, Yoo2002, Tanikawa2003, Wells2006, Dangelo2009}. A popular description is that electrons, tunneling from a STM tip onto the surface, propagate along the metallic surface-state before entering the bulk. To estimate the extension of a voltage drop on a surface with and without a metallic surface-state, we used finite element analysis (FEA) software to illustrate these two cases \cite{Comsol2011}.
\begin{figure}
\includegraphics{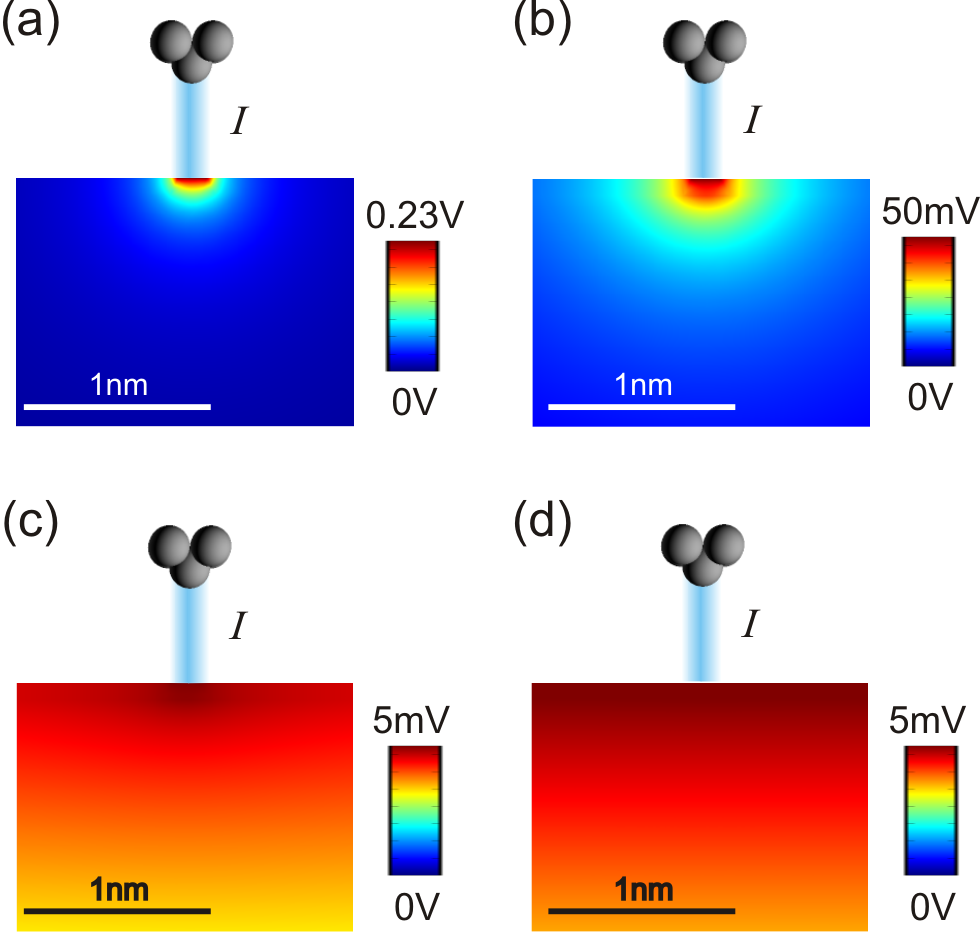}
\caption{\label{fig:PRBAbb1}(Color online) Finite element analysis of a voltage drop in a plain bulk material a) and a bulk material covered with a thin metal sheet on top b)\,-\,d) are shown. The bulk material has a constant conductivity of $\sigma_b$\,=\,10$\frac{\mathrm{S}}{\mathrm{m}}$ (equal to the Si(111) sample used in the experiment). In b)\,-\,d) the metallic surface-state of Si(111)-7$\times$7 is modelled by a metal sheet with a thickness of 100\,pm. The conductivities of the metal sheet increase with $\sigma_s$\,=\,10$^2\,\frac{\mathrm{S}}{\mathrm{m}}$, b), $\sigma_s$\,=\,10$^4\,\frac{\mathrm{S}}{\mathrm{m}}$, c), to $\sigma_s$\,=\,10$^6\,\frac{\mathrm{S}}{\mathrm{m}}$ in d). Without the conductive surface layer, a), the voltage drop amounts to 230\,mV and is highly localized, while the conductive surface layer leads to a reduction and a lateral spreading of the voltage drop.}
\end{figure}
For a surface without a metallic surface-state, we modelled a silicon semiconductor sample with a constant conductivity, shown in Fig.\,\ref{fig:PRBAbb1}\,a). To mimic a surface with a metallic surface-state, we added a thin metal sheet on top of the sample (Fig.\,\ref{fig:PRBAbb1}\,b)\,-\,d)). The FEA was performed with increasing conductivities of the metal sheet, $\sigma_s$\,=\,10$^2\,\frac{\mathrm{S}}{\mathrm{m}}$, b), $\sigma_s$\,=\,10$^4\,\frac{\mathrm{S}}{\mathrm{m}}$, c), and $\sigma_s$\,=\,10$^6\,\frac{\mathrm{S}}{\mathrm{m}}$, d), since these metal sheet conductivites cover the range of surface-state conductivities noted in Ref.\cite{Dangelo2009}, with e.g. 10$^4$\,$\frac{\mathrm{S}}{\mathrm{m}}$\,$\cdot$\,1\,\AA\; corresponding to 1\,$\frac{\mathrm{\mu S}}{\mathrm{\Box}}$. In each case, a)\,-\,d), we defined a highly localized current source on the surface. The current density was set to $\frac{1\,\mathrm{nA}}{(100\,\mathrm{pm})^2\pi}\,=\,$31\,$\frac{\mathrm{GA}}{\mathrm{m}^2}$. In Fig.\,\ref{fig:PRBAbb1}\,a), without a metal sheet, the voltage drop amounts to 230\,mV and is highly focused. In Fig.\,\ref{fig:PRBAbb1}\,b)\,-\,d), with a metal sheet, the voltage drop shows a reduction of its amount and an increasing lateral extension for increasing conductivities of the metal sheet. As we have observed the phantom force effect on the Si$($111$)$-7$\times$7 surface, the question remains how the metallic surface-state relates to the phantom force.\\
This article gives a description of the phantom force based on our model of an attractive electrostatic force in section \ref{sec:Theory}. Section \ref{sec:ExpMethods} introduces to the equipment and methods used for the experiments.
Following the conclusions from our finite element analysis, the phantom force is expected to occur on a surface without a metallic surface-state. This has not yet been demonstrated experimentally. In section \ref{sec:Hydrogenated}, we show the phantom force effect on a sample that does not have a metallic surface-state. In Section \ref{sec:Macroscopic} we investigated if the metallic surface-state has an effect on our observations on Si$($111$)$-7$\times$7. If the metallic surface-state would establish a constant potential over the whole surface even directly beneath the tip, we would expect a delocalized effect and thus, we would expect the observed phantom force to depend on the macroscopic tip radius $R$, just as the electrostatic force between a plate and a semi-spherical tip depends on the radius \cite{Olsson1998,Hudlet1998}. To evaluate this, we performed
constant height images on Si$($111$)$-7$\times$7 to extract the ratio between the frequency shift due to the phantom force and the tunneling current (\lq phantom force slope\rq, defined in eq. \ref{eq:PFseq}) and relate it to the macroscopic tip radius $R$, which was determined by force versus distance spectroscopy at zero effective bias.\\

\section{Theoretical description of the phantom force \label{sec:Theory}}

In this section, we introduce FM-AFM and describe the expected relation between the FM-AFM signal and the tunneling current in contrast to the relation between FM-AFM signal and tunneling current with a phantom force. Additionally, we mathematically derive the contribution of the tunneling current on the electrostatic force.\\
\begin{figure}
\includegraphics{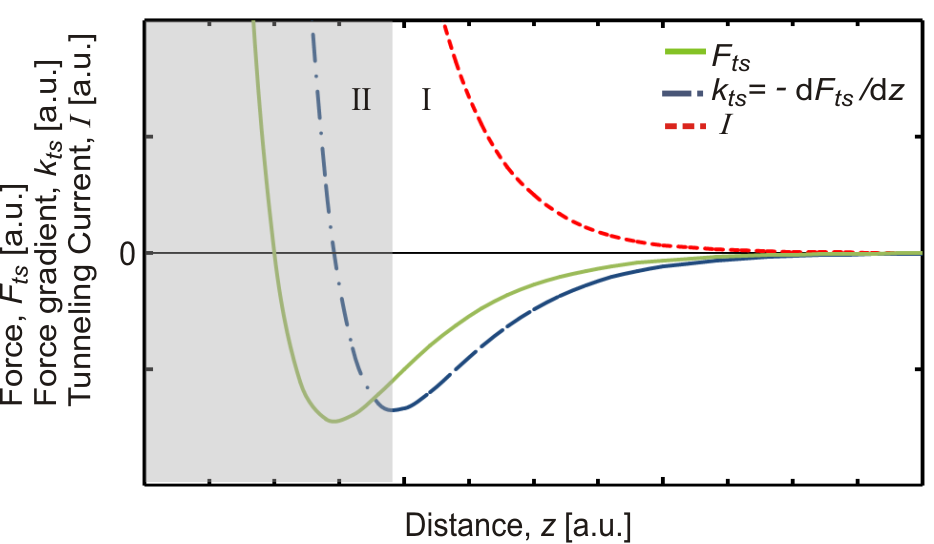}
\caption{\label{fig:PRBAbb2}(Color online) Qualitative distance dependence of the force $F_{ts}$ according to
a Morse potential, the corresponding force gradient $k_{ts}$ and
the tunneling current $I$. For small cantilever amplitudes, the
frequency shift $\Delta f$ is proportional to $k_{ts}$. In region
I, $\Delta f$ decreases, whereas $I$ increases. In region II,
$\Delta f$ starts to increase. }
\end{figure}
In FM-AFM, the forces between tip and sample cause a frequency shift $\Delta f$ relative to the resonance
frequency $f_0$ of an unperturbated cantilever \cite{Albrecht1991,Giessibl2003}. The cantilever has a stiffness
$k$ and oscillates with a constant amplitude $A$ at a distance $z$ from the surface. For small amplitudes, the
relation between the force $F_{ts}$ and $\Delta f$ is $\Delta f \approx \frac{f_0}{2 \cdot k}\cdot k_{ts}$,
where $k_{ts}=-\frac{d F_{ts}}{dz}$ is the force gradient between tip and sample \cite{Giessibl2001}. For a Morse
potential, which describes the chemical interaction between tip and sample atom, the force and $\Delta f$ behave
as shown in Fig.\,\ref{fig:PRBAbb2}. In region I, as the tip approaches the sample, the force becomes more attractive
and $\Delta f$ more negative. Approaching the tip closer to the sample, attractive chemical bonds start to form and
$\Delta f$ decreases further. In region II, $\Delta f$ starts to increase. In this article, we focus on region I,
which is the region at the onset of current. A more detailed explanation of the behaviour between force and $\Delta f$
is given e.g. in Ref.\cite{Giessibl2003}. When performing STM, the tunneling current exponentially increases with
decreasing tip-sample distance \cite{Binnig1983}. If force and current were independent, we would expect, a decrease in the frequency shift as the tunneling current increases on a $\Delta f$ versus $I$ plot.\\
However, a surprising characteristic of the phantom force is the increase of the frequency shift as the tunneling
current increases when plotting $\Delta f$ against $I$ \cite{Weymouth2011}.\\
The phantom force can be modelled by extending the formula of the attractive electrostatic force between two metal
objects by a tunneling current dependent term. Without the influence of the tunneling current we can write
\begin{equation}
F_{ts}^{es}\,=\,-\,\frac{1}{2}\,\frac{dC_{ts}}{dz}\,V^2\ ,
\label{eq:wideeq}
\end{equation}
where $V$ is the potential difference between tip and $C_{ts}$
is the capacity of the tip-sample junction. If the tip was a flat
surface $A$ at a distance $z$ to the sample, the derivative of
capacity with distance would be given by
\begin{equation}
\frac{dC_{ts}}{dz} = - \epsilon_0 \frac{A}{z^2} \label{eq:dC_ts}.
\end{equation}
The permittivity of vacuum $\epsilon_0 \approx 8.85$\,pF/m can also be expressed as $\epsilon_0 \approx 8.85$\,pN/V$^2$. Thus, for
$A=20\,z^2$, a force of about 90\,pN would arise for a bias of 1 Volt, increasing with the square of voltage.\\
The effective bias responsible for the electrostatic force is $V=V_{bias}\,-\,V_{CPD}$, where $V_{bias}$ is the
applied voltage and $V_{CPD}$ is the contact potential difference between tip and
sample \cite{Nonnenmacher1991}. While local changes in $V_{CPD}$  \cite{Sadewasser2009}
will affect the local electrostatic attraction between tip and sample dependent upon $V_{bias}$,
they cannot explain observations of this phantom force, for reasons discussed in Ref.\,\cite{Weymouth2011}:
A local change in $V_{CPD}$ would cause a $\Delta f$ decrease in one bias (assuming the applied $|V_{bias}|>|V_{CPD}|$)
and an increase in the opposite bias, whereas we observe an increase in $\Delta f$ with significant bias independent of polarity.\\
We thus consider the voltage $V$ being modified by a voltage drop caused by the
tunneling current passing through the sample with resistivity $R_s$. Therefore, $V\,=\,V_{bias}\,-\,I\cdot R_s$. The electrostatic force is then
\begin{equation}
F_{ts}^{es}=-\frac{1}{2}\frac{dC_{ts}}{dz}(V_{bias}^2\,-\,2\,V_{bias}\,I\,R_s\,+\,I^2\,R_s^2)
\label{eq:shorteq}
\end{equation}
with an offset component proportional to $V_{bias}^2$, a term linear with $I$ and
a quadratic term in $I$. At typical experimental conditions as in our previous experiments, where
$V_{bias}$\,=\,1.5\,V, $R_s$\,=\,150\,M$\Omega$ and $I$\,=\,1\,nA
\cite{Weymouth2011}, the quadratic term is 5\,\% of the linear term and can be neglected (however, for
very small tip-sample distances as required for atomically resolved AFM on low-conductivity samples, this term can not be neglected). Without the quadratic term,
equation \ref{eq:shorteq} reduces to
\begin{equation}
F_{ts}^{es} \approx -\frac{1}{2}\frac{dC_{ts}}{dz}(V_{bias}^2 - 2 V_{bias} I R_s ).
\label{eq:simplyeq}
\end{equation}
In order to determine a relation between the frequency shift $\Delta f$ and the tunneling current $I$,
we first have to calculate the contribution of this electrostatic force to the force gradient,
$k_{ts}^{es}$. After substituting $I\,=\,I_0\,e^{-\kappa z}$ into equation \ref{eq:simplyeq} and taking the
derivative of $F_{ts}^{es}$ with respect to $z$, equation \ref{eq:simplyeq} results in
\begin{equation}
k_{ts}^{es}\,=\frac{1}{2}\frac{d^2C_{ts}}{dz^2}\,V_{bias}^2\,-\,\left(\frac{d^2C_{ts}}{dz^2}-\frac{dC_{ts}}{dz}\kappa \right)\,V_{bias}\,I\,R_s\ .
\label{eq:stiffeq}
\end{equation}
Since $\Delta f$ is directly proportional to $k_{ts}^{es}$, assuming the small amplitude approximation, equation \ref{eq:stiffeq} can be rewritten as
\begin{equation}
\Delta f\,=\frac{f_0}{4k}\frac{d^2C_{ts}}{dz^2}\,V_{bias}^2 -\frac{f_0}{2k}\,\left(\frac{d^2C_{ts}}{dz^2}-\frac{dC_{ts}}{dz}\kappa\right) V_{bias} R_s I
\label{eq:dfeq}
\end{equation}
The $\Delta f$ line shows a linear dependence with $I$ with an offset that depends on capacity and bias and a slope that is linear with $R_s$ and $V_{bias}$. We define, from equation \ref{eq:dfeq}, the phantom force slope as
\begin{equation}
\Xi\,:=\,\frac{d (\Delta f)}{d I}\,=\,-\frac{f_0}{2k}\,\left(\frac{d^2C_{ts}}{dz^2}-\frac{dC_{ts}}{dz}\kappa\right)\,V_{bias}\,R_s\ ,
\label{eq:PFseq}
\end{equation}
which is usually expressed in $\frac{\mathrm{Hz}}{\mathrm{nA}}$. The slope is a measure of the strength of the phantom force.\\

\section{Experimental methods and setup\label{sec:ExpMethods}}

The experiments were performed in ultrahigh vaccum ($\approx\,3\cdot10^{-10}$\,Torr) and at room temperature.
The images in this article were all aquired in constant height mode. qPlus sensors ($k$\,=\,1800\,$\frac{\mathrm{N}}{\mathrm{m}}$)
were equipped with tungsten tips to probe the sample. The tungsten tips were prepared by common techniques like controlled
collision with the sample, field emission and explosive delamination \cite{Hofmann2010}.\\
The n-doped Si(100) samples had a resistivity of 0.008\,-\,0.012\,$\Omega$cm at 300\,K. The surface was prepared by repeated cycles
of flashes up to 1250\,$^{\circ}$C followed by cooling periods in the range of several minutes. After cleaning, approximately 300\,L
deuterium \cite{deposition} were deposited onto the surface at $\approx$450\,$^{\circ}$C.\\
The Si$($111$)$-7$\times$7 samples used were p-doped with $\rho$\,$=$\,6\,-\,9\,$\Omega$cm at 300\,K. The surface was cleaned by the
same flash routine as described above.\\

\textbf{Investigation of a potential preamplifier artifact}\\

\begin{figure}
\includegraphics{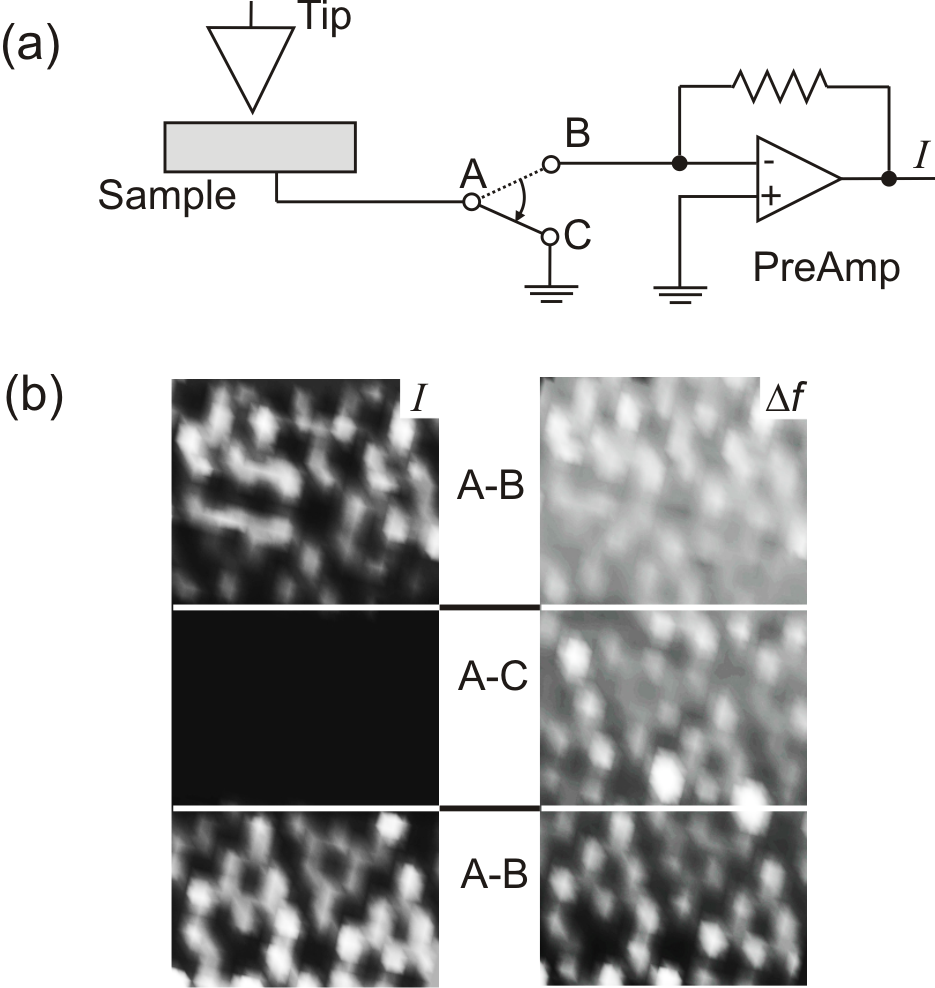}
\caption{\label{fig:PRBAbb3} a) Schematic of
the experimental setup. With an implemented switch, the preamp can
be used for the $I$ signal (switch in position A\,-\,B) or can be
shorted to ground (switch in position A\,-\,C). In b),
simultaneous aquired $I$ and $\Delta f$ data in constant height
mode are shown. The images were scanned from top to bottom. In the
middle section the $I$ signal was shorted to ground to check if the
phantom force effect is real (i.e. caused by a voltage drop at the sample surface) or caused by a preamp artifact.}
\end{figure}

In the experimental setup, the bias voltage $V_{bias}$ was applied
to the tip with the sample referenced to virtual ground via a
preamplifier (\lq preamp\rq). The preamp is attached outside
vacuum and amplifies, as a current-to-voltage converter, the $I$
signal by a factor of 10$^8$\,$\frac{\mathrm{V}}{\mathrm{A}}$.
Since the tip is oscillating, $I$ is an alternating current (AC)
with a direct current (DC) offset, where only the DC component is
measured by the preamp due to its limited bandwidth. To
investigate if this phantom force effect is not due to a
fluctuation of the virtual ground of the preamp, we introduced a
switch that allows to either connect the sample to real ground via
a direct ground connection or the virtual ground of the preamp, as
schematically shown in Fig.\,\ref{fig:PRBAbb3}\,a). Because the
operational amplifier used in the preamp has a limited gain,
limited bandwidth and a limited slew rate (in contrast to an ideal
operational amplifier), the virtual ground terminal can deviate
from zero, and cross-coupling to the force gradient measurement might occur.

In the upper and lower section of Fig.\,\ref{fig:PRBAbb3}\,b),
simultaneously recorded $I$ and $\Delta f$ data are presented in
constant height mode (switch in position A\,-\,B). In the middle
section of the image, the $I$ signal from the sample was shorted
to ground (switch in position A\,-\,C). Nevertheless, the phantom force effect is still present in the $\Delta f$ signal, which
clearly demonstrates that the phantom force is not caused by a
preamp artifact, but by the current-induced local potential deviation outlined
above.\\

\section{The phantom force on the hydrogenated Si(100) surface \label{sec:Hydrogenated}}

In this section, we present measurements on the H/Si(100) surface. Exposing Si(100) to hydrogen
saturates the unsaturated dangling bonds \cite{Boland1990}. The electronic states of the hydrogenated
dimers have been shown to be outside the bandgap of bulk Si, meaning that in contrast to Si(111)-7$\times$7,
the surface does not have a metallic surface-state \cite{Raza2007}.\\
\begin{figure}
\includegraphics{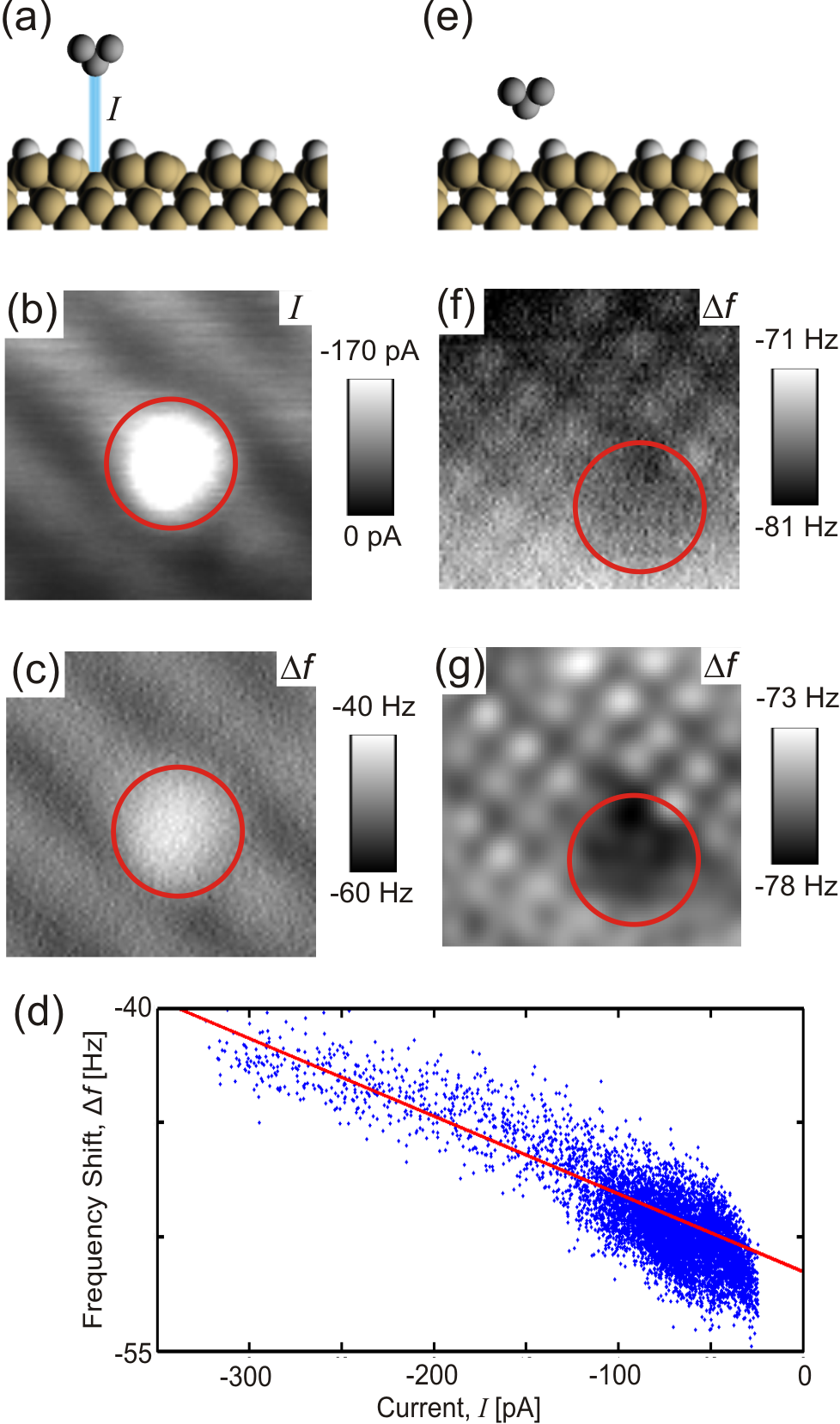}
\caption{\label{fig:PRBAbb4}(Color online) a) shows a tunneling current, which induces the phantom force effect. Simultaneous acquired $I$ and $\Delta f$ data with atomic contrast are shown in b) and c) for $V_{bias}$\,$=$\,1.5\,V. In d), the relation between $\Delta f$ and $I$ data is plotted. Data points with increased $I$ data, as the defect outlined in red, show a stronger decrease (less attraction) of the $\Delta f$ values. However, at close tip-sample distances, e), and low bias, in this case 200 mV, the defect appears darker (more attractive), f). Image g) shows f) with low-pass filtering and plane substraction applied for clarity. Images are 2\,nm$\times$2\,nm, $A$\,=\,100\,pm, $k$\,=\,1800\,$\frac{\mathrm{N}}{\mathrm{m}}$, $f_0$\,=\,19131\,Hz.}
\end{figure}
In Fig.\,\ref{fig:PRBAbb4}\,a), a tunneling current between tip and sample, which leads to the phantom force effect, is schematized.
Fig.\,\ref{fig:PRBAbb4}\,b) and c) show simultaneous $\Delta f$ and $I$ data collected at constant height with an applied bias
voltage of 1.5\,V. The dimer rows can be seen running from upper left to lower right. The low contrast is due to our choice of a relatively
large imaging distance to prevent excessive tunneling currents when scanning over the defect area, circled in red. The feature circled in red is most likely a dangling bond, which we would expect to observe in $\Delta f$ data as darker (more attractive). However due to the increase of the tunneling current over it, the phantom force effect causes an increase in $\Delta f$ that makes it appear brighter. To investigate the relationship between $\Delta f$ and $I$, we plotted in Fig.\,\ref{fig:PRBAbb4}\,d) the $\Delta f$ information of each single pixel in image c) versus the corresponding pixel of the $I$ information in b). For positive applied bias voltages, the $I$ signal is negative. The relation between $\Delta f$ and $I$ data results in $\Xi$\,=\,-\,34$\frac{\mathrm{Hz}}{\mathrm{nA}}$, if we assume a linear relation as described in equation \ref{eq:dfeq}.\\
In Fig.\,\ref{fig:PRBAbb4}\,e), the bias voltage is decreased to 200\,mV and in order to resolve atomic contrast, the tip must be approached to the surface, similar to our previous observations of the phantom force \cite{Weymouth2011}. The attractive interaction in the presence of the dangling bond is clearly observed in $\Delta f$ data collected at low bias, as shown in Fig.\ref{fig:PRBAbb4}\,f). Fig.\ref{fig:PRBAbb4}\,g) is a low-pass filtered and plane substracted image from f) to show the dangling bond with better contrast.\\
We demonstrated that the phantom force does not depend on the presence of a metallic surface-state and still appears on a sample system as H/Si(100) without a metallic surface-state.\\

\section{The dependence of the phantom force on the macroscopic tip radius on Si(111)-7$\times$7 \label{sec:Macroscopic}}

\begin{figure}
\includegraphics{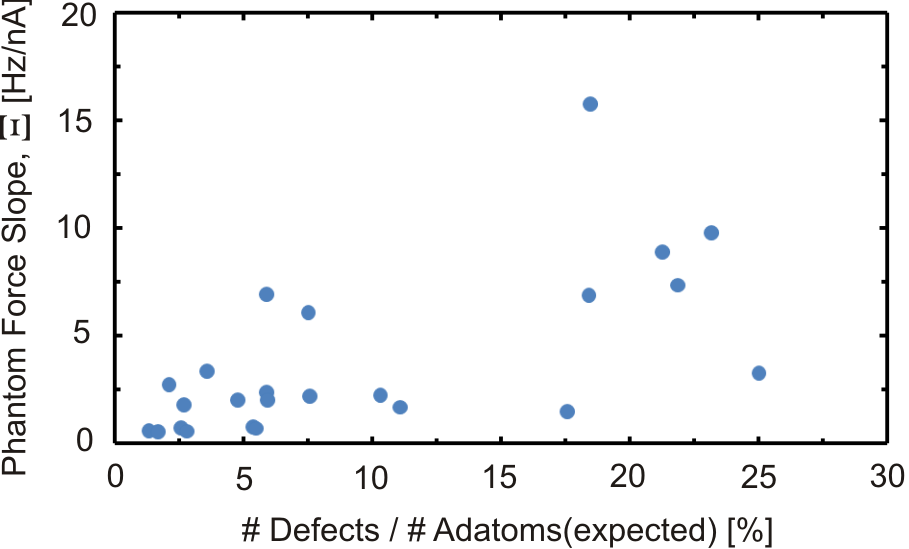}
\caption{\label{fig:PRBAbb8}(Color online) phantom force slopes $\Xi$ plotted versus the defect density on the Si(111)-7$\times$7 surface. The phantom force slopes  $\Xi$ seem small for less defects on the Si(111)-7$\times$7 and more defects point to increased $\Xi$. }
\end{figure}

In the following section, we investigate the phantom force on the Si(111)-7$\times$7 surface. If the metallic surface-state plays a role we would expect a delocalized effect. Then we should observe a pronounced dependence of the phantom force as a function of the macroscopic tip radius $R$ \cite{Olsson1998,Hudlet1998}. We also discuss the results based on the factors of the phantom force slope $\Xi$ introduced in section \ref{sec:Theory}.\\
\begin{figure}
\includegraphics{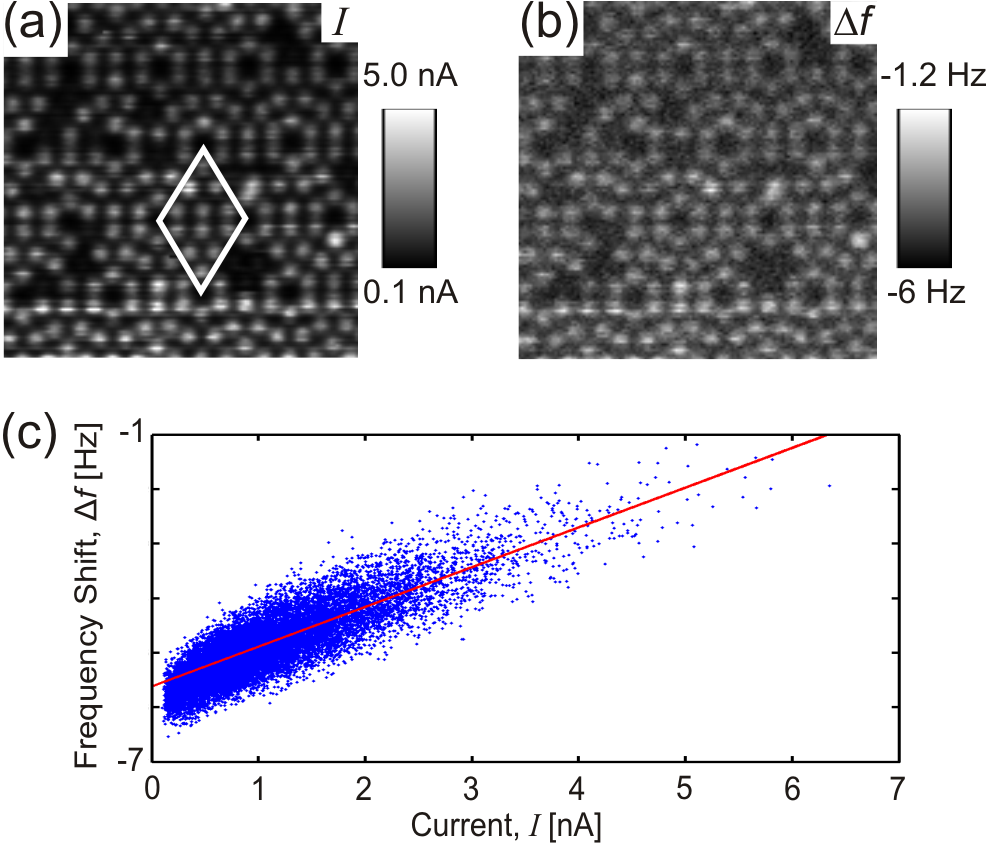}
\caption{\label{fig:PRBAbb5}(Color online) a) Tunneling current $I$ and b) $\Delta f$ data taken during constant height scanning. A Si$($111$)$-7$\times$7 unit cell is outlined by a white diamond in a). In b) the adatoms appear more bright, which is due to less attraction induced by the tunneling current. c) shows the phantom force contribution as a linear dependence between the current and $\Delta f$ data. The phantom force slope $\Xi$ of the line is 0.73\,$\frac{\mathrm{Hz}}{\mathrm{nA}}$. Images are 10\,nm$\times$10\,nm, $A\,\approx$\,400\,pm, $k$\,=\,1800\,$\frac{\mathrm{N}}{\mathrm{m}}$, $f_0$\,=\,19\,130\,Hz and $V_{bias}$\,$=$\,-1.5\,V.}
\end{figure}
Figure \ref{fig:PRBAbb5}\,a) and b) show simultaneously acquired $I$ and $\Delta f$ data of the Si$($111$)$-7$\times$7 surface. In Fig.\,\ref{fig:PRBAbb5}\,a) the tunneling current reaches its maximium above the adatoms. In Fig.\,\ref{fig:PRBAbb5}\,b) the brighter adatoms show a repulsive force contribution. The frequency shift is less negative over regions with a high tunneling current. A linear dependence between $\Delta f$ and the $I$ signal is shown in Fig.\,\ref{fig:PRBAbb5}\,c). By fitting the data we extracted a phantom force slope $\Xi$\,=\,0.73\,$\frac{\mathrm{Hz}}{\mathrm{nA}}$.\\
\begin{figure}
\includegraphics{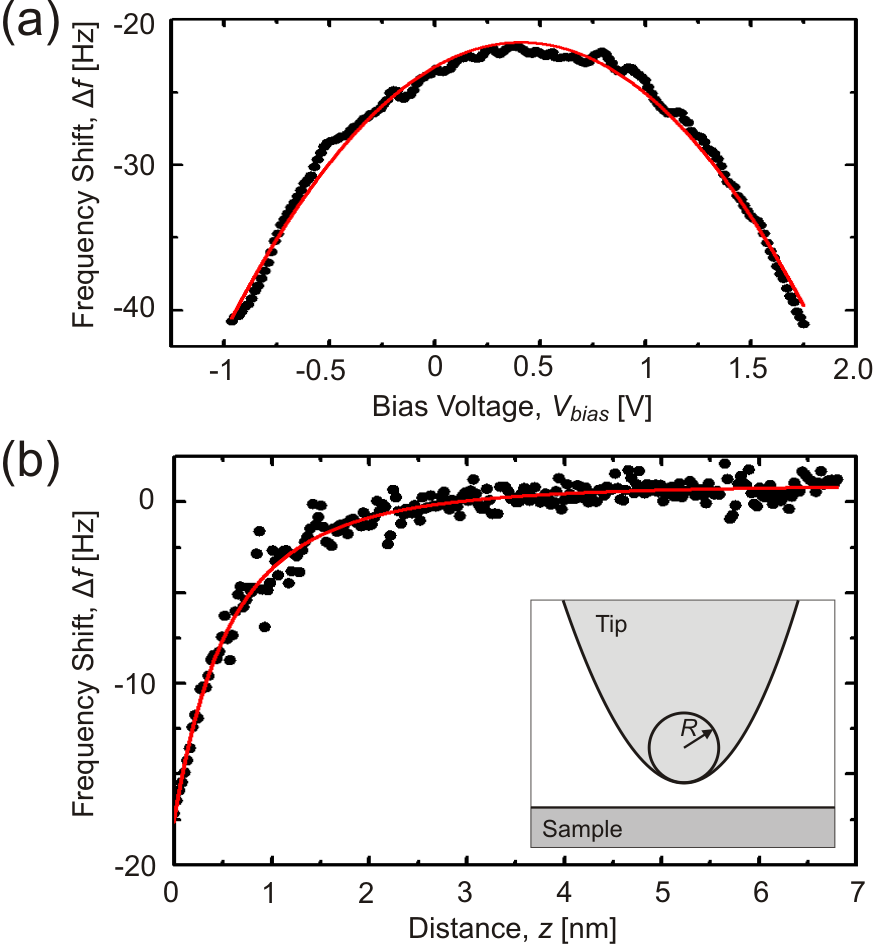}
\caption{\label{fig:PRBAbb6}(Color online) a) Spectrum of $\Delta f(V_{\mathrm{bias}})$ for determining the $V_{CPD}$ between tip and sample to 0.4\,V. b) Spectrum of $\Delta f(z)$ taken at $V_{CPD}$. The macroscopic tip radius $R$ was extracted by fitting the curve to a long range van der Waals force contribution, for b) the extracted $R$\,=\,600\,nm. The inset shows the tip radius for a parabolic tip shape.}
\end{figure}
The macroscopic tip can be described by its tip radius $R$, which we determined by fitting the long-range $\Delta f$ contribution between tip and sample to a model assuming only van der Waals interaction \cite{Hoelscher1999}. In order to minimize the attractive electrostatic force, we took $\Delta f(z)$ spectra while compensating for the V$_{CPD}$. Before measuring the $\Delta f(V_{\mathrm{bias}})$, we retracted the tip from the sample by 100\,pm. This reduced the possibility of tip-sample collisions due to drift. The voltage corresponding to the maximum $\Delta f$ value of the parabolic $\Delta f(V_{\mathrm{bias}})$ curve equals to $V_{CPD}$ \cite{Guggisberg2000}. In Fig.\ref{fig:PRBAbb6}\,a) the $V_{CPD}$ was determined to 0.4\,V. The $\Delta f(z)$ curves were fitted to a model incorporating a parabolic tip shape (as shown in Fig.\ref{fig:PRBAbb6}\,b)) in accordance to Refs.\cite{Hoelscher1999,Giessibl1997}. The fit of the $\Delta f(z)$ data in Fig.\ref{fig:PRBAbb6}\,b) result in $R$\,=\,600\,nm.\\
\begin{figure}
\includegraphics{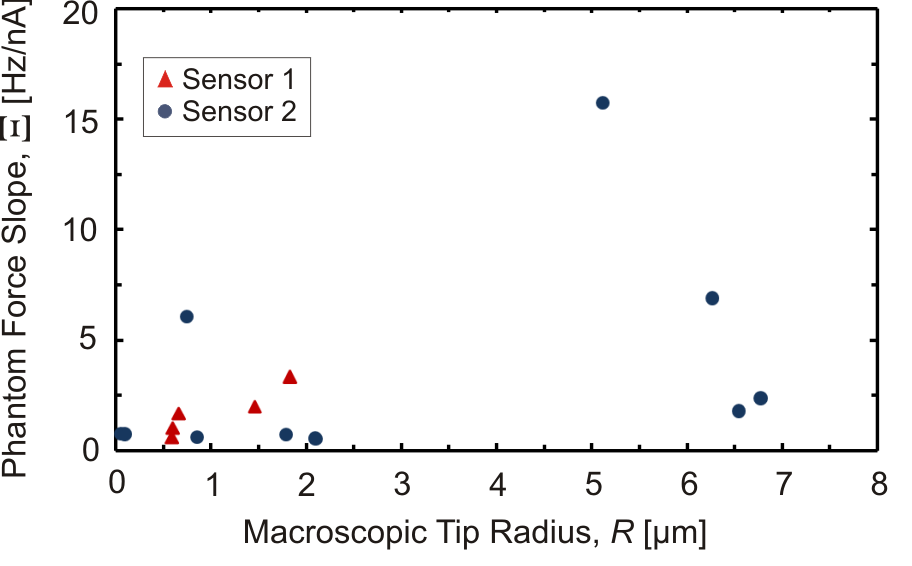}
\caption{\label{fig:PRBAbb7}(Color online) The phantom force slope $\Xi$ in $\frac{\mathrm{Hz}}{\mathrm{nA}}$ characterizing the phantom force is plotted versus the macroscopic tip radii $R$. Two different sensors were used, sensor\,1 (red triangles) and sensor\,2 (blue dots). The phantom force slopes $\Xi$ show no dependence on the macroscopic tip radii $R$.}
\end{figure}
Fig.\,\ref{fig:PRBAbb7} displays different phantom force slopes $\Xi$ dependent on the respective macroscopic tip radius $R$. Sixteen data points, acquired with two different qPlus sensors (sensor 1:\,red triangles and sensor 2:\,blue dots), are plotted. The data points are widely spread and range from radii of 51\,nm to 6775\,nm. The phantom force slopes vary from 0.51\,$\frac{\mathrm{Hz}}{\mathrm{nA}}$ to 15.74\,$\frac{\mathrm{Hz}}{\mathrm{nA}}$. In particular, slopes below 2.0\,$\frac{\mathrm{Hz}}{\mathrm{nA}}$ can be observed for a wide range of macroscopic tip radii $R$. We observe no dependence between the phantom force slope and the macroscopic tip radius. This supports our hypothesis of a highly local voltage drop, and suggests that the metallic surface-state does not play a role.\\
The spread in our measurements of Fig.\,\ref{fig:PRBAbb7} ($\Xi$\,$>$\,5.0\,$\frac{\mathrm{Hz}}{\mathrm{nA}}$) can be discussed with the aid of equation \ref{eq:PFseq}, $\Xi=-\frac{f_0}{2k}(\frac{d^2C_{ts}}{dz^2}-\frac{dC_{ts}}{dz}\kappa) V_{bias} R_s$. We turn now to the factors $-\frac{f_0}{2k}(\frac{d^2C_{ts}}{dz^2}-\frac{dC_{ts}}{dz}\kappa)$ and $R_s$ in detail, as $V_{bias}$ was constant for these measurements.\\
The factor $-\frac{f_0}{2k}(\frac{d^2C_{ts}}{dz^2}-\frac{dC_{ts}}{dz}\kappa)$ can be calculated, assuming a model of the electrostatic force $F_{ts}^{es}$ of a conical tip (half-angle $\theta$) in front of a metallic surface as described by Hudlet \textit{et al.} \cite{Hudlet1998}. This would be applicable, if the tip and sample surfaces could be modelled by a constant potential. Calculations for realistic $R$=\,5\,nm and $\theta$=\,70$^{\circ}$, at conditions summarized in Ref.\cite{PhantomSlope}, lead to unrealistic values of $\Xi$= 68$\frac{\mathrm{Hz}}{\mathrm{nA}}$, much larger than $2.8\,\frac{\mathrm{Hz}}{\mathrm{nA}}$, the experimentally determined average $\Xi$ of the values shown in Fig.\,\ref{fig:PRBAbb7}. We propose that this phantom force effect is highly localized. Instead of being described by the macroscopic tip shape, it would be better described by a model of the nanoscopic tip cluster. Assuming a plate capacitor with $C = \epsilon_0 \frac{A}{z}$, we can calculate the phantom force slope using equation \ref{eq:PFseq} and the following parameters: $f_0=20$\,kHz, $k=1800$\,N/m, $\kappa=2$\,\AA$^{-1}$, $R_s=150$\,M$\Omega$ with an applied bias $V_{bias}= -1.5$\,V, at a distance $z=4.4$\,\AA\ and a capacitive area $A=(1$\,nm$)^2$. This yields a slope $\Xi=2.8$\,Hz/nA, equivalent to the experimental average of $2.8\,\frac{\mathrm{Hz}}{\mathrm{nA}}$.\\
Concerning the factor $R_s$, we observed in Ref.\,\cite{Weymouth2011} that the higher the sample resistivity the higher the slope $\Xi$. In our case, the data points with higher $\Xi$ were collected on areas with an increased number of defects on the Si(111)-7$\times$7 surface. The dependence between $\Xi$ and the defect density on the Si(111)-7$\times$7 surface was investigated and is shown in Fig.\ref{fig:PRBAbb8}. For low defect densities, $\Xi$ seems to be low in contrast to higher defect densities with an increased $\Xi$. But, since the tip shape and $R_s$ changed for each data point, the dependence between phantom force slopes and the defect densities is not conclusive and has to be investigated in more detail.

\section{Summary and Outlook}

We showed in section \ref{sec:Hydrogenated} that the phantom force is present on a sample system without a metallic surface-state.\\
In section \ref{sec:Macroscopic}, we investigated the influence of a metallic surface-state on the phantom force.
The experimental observation of the phantom force slope $\Xi$ shows no dependence on the macroscopic tip radius $R$. This infers a highly localized voltage drop and we concluded that the metallic surface-state does not play a role for the phantom force effect.\\
For a future project we suggest low temperature measurements to investigate the dependence of the phantom force on the defect density on Si(111)-7$\times$7. In this experiment, the tip would be more stable and a controlled exposure of a distinct spot on the surface to e.g. oxygen could clarify the dependence between phantom force and sample resistivity.

\section*{Acknowledgments}

The authors thank the German Science Foundation (DFG,
Sonderforschungsbereich 689) for financial support, J. Welker, M.
Emmrich, E. Wutscher and F. Pielmeier for helpful discussions.


\begin{thebibliography}{99}
\bibitem{deposition}The Si(100) surface was saturated with deuterium, which for purposes of this study, behaves as hydrogen.
\end{thebibliography}

\end{document}